\documentclass[aps,twocolumn,prb,showpacs,superscriptaddress,floatfix]{revtex4}

\usepackage{graphics,amssymb,amsmath,times}

\begin{document}

\title{Quantum and classical diffusion in small-world networks}

\author{Beom Jun \surname{Kim}}
\email{beomjun@ajou.ac.kr}
\affiliation{Department of Molecular Science and Technology, 
  Ajou University, Suwon 442-749, Korea}
\author{H. \surname{Hong}}
\affiliation{School of Physics, Korea Institute for Advanced Study, Seoul 130-722, Korea}
\author{M.Y. \surname{Choi}}
\affiliation{Department of Physics, Seoul National University, Seoul 151-747, Korea}
\affiliation{School of Physics, Korea Institute for Advanced Study, Seoul 130-722, Korea}

\begin{abstract}
We study numerically quantum diffusion of a particle on small-world networks
by integrating the time-dependent Schr\"odinger equation with a localized 
initial state.  
The  participation ratio, which corresponds to the number of visited sites 
in the case of classical diffusion,  as a function of time is measured and the
corresponding diffusion time $\tau$ is computed.  In a
local regular network, i.e., in the network with the rewiring probability $p=0$, 
the diffusion time depends on the network size $N$ as $\tau \sim N$, 
while the behavior $\tau \sim \log N$ 
is observed as $p$ becomes finite.  Such fast diffusion of a particle on
a complex network suggests that the small-world transition
is also the fast-world transition from a dynamic point of view.
The classical diffusion behavior is also studied and compared with the
quantum behavior.
\end{abstract}


\maketitle

%
There has been a surge of research activity on various aspects
of complex networks since some important features of 
real networks were successfully explained by 
simple model networks.~\cite{ref:review,ref:WS,ref:BA}
In particular, the Watts-Strogatz (WS) model~\cite{ref:WS}
was the first to produce networks with small-world behavior,
characterized by the path length increasing logarithmically 
with the network size.
%
Subsequently, a number of studies have been performed on such complex 
networks, focused mostly on structural properties of the networks.
On the other hand, vertices of a real network may have some internal 
degrees of freedom, interwoven with the structure of the network. 
Motivated by this observation, a group of papers have also investigated 
statistical mechanical models defined on complex networks.
For example, spin models like the Ising model and the $XY$ model 
on WS networks have been shown to undergo finite-temperature phase 
transitions of the mean-field nature,~\cite{ref:smallIsing,ref:XY} 
disclosing the role of long-range interactions along the shortcuts. 
In some cases the time-development of the network structure may be 
coupled to the dynamics of degrees of freedom defined on vertices. 
Still the study of vertex dynamics without considering its influence on 
the network structure can be useful as a first step toward the 
complete understanding.  In this spirit, dynamic models defined on networks 
draw much attention: Epidemic spreading as well as classical diffusion on 
complex networks have been studied,~\cite{ref:diff,ref:epidemic} and 
very recently, the dynamic universality class of the $XY$ model on the WS 
network has been identified.~\cite{ref:dynamicXY}

Properties of a quantum mechanical model put on a complex network have 
also been studied with respect to the spectral properties of the Laplacian 
operator \cite{ref:spec} and the localization-delocalization
transition in the presence of site disorder.~\cite{ref:quantum}
In this work we consider quantum as well as classical diffusion
of a particle on the WS network without site disorder, and investigate
how the diffusion time scales with the network size. 
It is well known that as soon as the WS network (of size $N$) has 
a finite fraction of shortcuts, it undergoes the small-world transition 
that the characteristic path length $l$ changes its behavior
from $l \sim N$ to $\log N$.~\cite{ref:WS,ref:Newman}
>From the dynamic point of view, this change from the large-world to the
small-world behavior is expected to be accompanied by a sharp change in 
the behavior of the diffusion time: If the world is small, the
traveling time around the world should also be short.
Indeed the diffusion time $\tau$ associated with the participation ratio
is observed to change the size dependence
from $\tau \sim N$ to $\log N$, which is to be compared with
the case of classical diffusion: $\tau \sim N^2$ to $N$.
%

%
The low-dimensional tight-binding electron system has been studied 
extensively in relation to the localization transition in
the presence of disorder. In the absence of disorder, the 
energy eigenstate of a tight-binding electron on a local 
regular lattice is always in the extended state due to the
Bloch theorem.  Here we consider the tight-binding electron on the
WS network, described by the time-dependent Schr\"odinger equation
\begin{equation} \label{eq:Schrodinger}
i \frac{ \partial | \Psi \rangle }{ \partial t} =  H | \Psi \rangle ,
\end{equation}
where we have set $\hbar \equiv 1$ and the ket $| \Psi \rangle$
has the position representation $\Psi_n \equiv \langle n | \Psi \rangle$
at the $n$th vertex. 
The Hamiltonian in the position representation takes the form
\begin{equation} \label{eq:H}
H_{nn^\prime} = H_{n^\prime n} = \left\{
\begin{array}{ll}
\Delta   &  \mbox{for $n^\prime \in \Lambda_n$} \\
0        &  \mbox{otherwise,} \\
\end{array}
\right.
\end{equation} 
where the on-site energy has been assumed to be uniform and 
set equal to zero ($H_{nn} \equiv 0$), $\Delta$ is the 
hopping energy, and $\Lambda_n$ represents the set of neighbors of 
vertex $n$.  For example, in a local regular network with the
connection range $r=1$, we have $\Lambda_n = \{ n{-}1,\, n{+}1 \}$. 

\begin{figure}
\centering{\resizebox*{!}{2.8cm}{\includegraphics{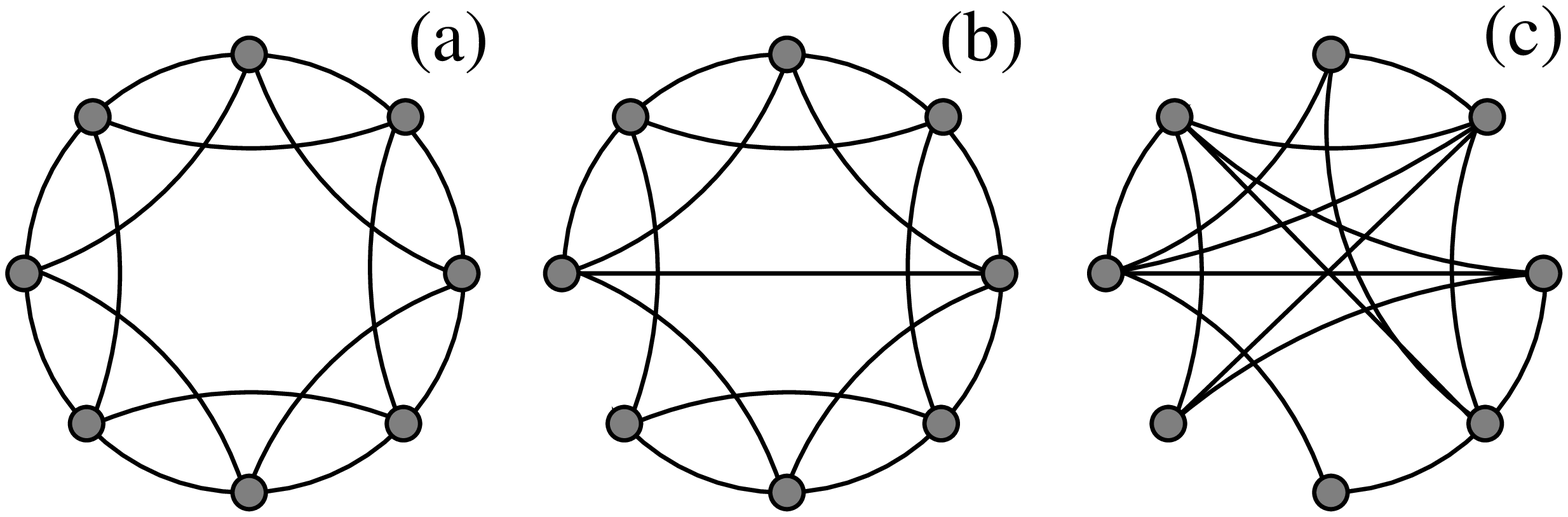}}}
\caption{
Construction of the Watts-Strogatz (WS) network.
Starting from the regular one-dimensional lattice in (a) with the
connection range $r=2$, we visit each edge and rewire it with
given probability $p$, thus generating shortcuts in (b).
Via this procedure a complicated network structure emerges, 
as shown in (c).
}
\label{fig:WS}
\end{figure}

The WS network is constructed according to the standard procedure in 
Ref.~\onlinecite{ref:WS}:
A one-dimensional (1D) local regular network with the connection range $r$
is built first, then with the rewiring probability $p$ each local
edge is rewired to a randomly chosen other vertex. (See Fig.~\ref{fig:WS},
illustrating the case $r=2$.)
Once the WS network is constructed in this way, we normalize the time $t$
in units of  $1/\Delta$, and integrate the time-dependent 
Schr\"odinger equation given by Eqs.~(\ref{eq:Schrodinger}) and
(\ref{eq:H}) numerically by means of the fourth order Runge-Kutta method 
with the discrete time step $\delta t = 0.01$, starting from the initial condition
that the electron is localized at randomly chosen vertex $m$, i.e.,
$\Psi_n (t{=}0) = \delta_{n, m}$.
For simplicity, we set the initial position $m\equiv 0$.

In the absence of shortcuts, the analytic solution of Eq.~(\ref{eq:Schrodinger})
is easily found:~\cite{ref:r1}
In the simplest case of $r=1$, the Fourier transformation 
${\widetilde \Psi_k} = \sum_n \Psi_n e^{i n k}$ yields
\begin{equation}
i \frac{ \partial {\widetilde \Psi_k} }{ \partial t}
= (e^{ik} + e^{-ik}){\widetilde \Psi_k} = 2\cos k {\widetilde \Psi_k} ,
\end{equation}
which in turn leads to 
\begin{equation}
{\widetilde \Psi_k}(t) = {\widetilde \Psi_k}(0) e^{-2it \cos k } .
\end{equation}
The inverse Fourier transformation 
\begin{equation}
\Psi_n = \frac{1}{N}\sum_k {\widetilde \Psi_k} e^{-ink}
\end{equation}
in the limit of $N \rightarrow \infty$, combined with the
Jacobi-Anger expansion
$e^{i z \cos\theta } = \sum_{m= -N}^N i^m J_m(z)e^{im\theta}$
gives the solution
\begin{equation} \label{eq:Jn}
|\Psi_n(t)| = |J_n(2t)|
\end{equation}
with the Bessel function $J_n$ of the order $n$.
For $r=2$, it is straightforward to obtain $\Psi_n(t) = 
\sum_p (-i)^{n-p} J_{n - 2p}(2t) J_p (2t)$.

\begin{figure}
\centering{\resizebox*{!}{6.0cm}{\includegraphics{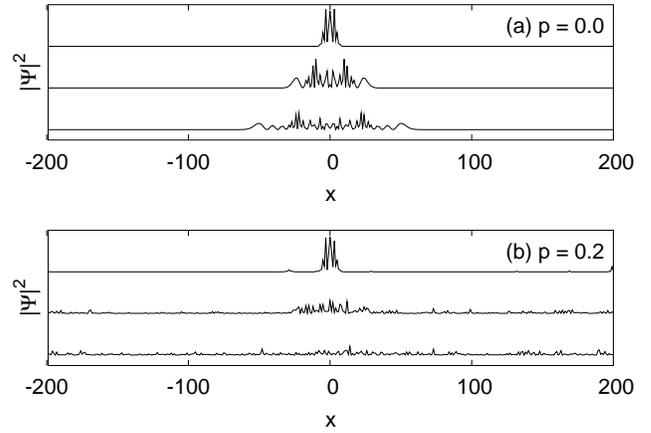}}}
\caption{
Spread of the wave packet initially localized at $x=0$ 
on the Watts-Strogatz network with the connection range $r=2$ 
for the rewiring probability $p=$ (a) $0$ and (b) $0.2$. 
In both (a) and (b), the three curves described the wave packet at 
time $t=1.0$, $5.0$, and $10.0$, respectively (from top to bottom). 
It is observed that the wave packet diffuses much faster in (b) 
than in (a).  For clarity, $2|\Psi|^2$ (instead of $|\Psi|^2$ 
for $t=1.0$) is plotted for $t=5.0$ and $10.0$. 
}
\label{fig:Psi}
\end{figure}

As time proceeds, the initial localized state diffuses and eventually
evolves to an extended state in which  $\Psi_n \neq 0$ at any $n$.
Figure~\ref{fig:Psi} shows such diffusion of the wave packet on
the WS network with the connection range $r=2$
for the rewiring probability $p= 0$ and $0.2$. 
It is clearly shown that diffusion occurs much faster in (b), 
namely, for $p \neq 0$. 
In this case as soon as the wave packet diffuses to a nearby end point 
of a shortcut, the other end of the shortcut becomes a new center for 
diffusion at the next time step, resulting in faster spread of the 
wave packet.  

To describe the diffusion of the wave packet, one may measure the
variance of the wave packet,~\cite{ref:GHE} $\sigma^2 (t) = 
\langle \Psi (t) | x^2 | \Psi (t) \rangle - \langle \Psi (t) | x 
| \Psi (t) \rangle^2 $ with the one-dimensional
position $x$ on the network. However, the variance $\sigma^2$
tends to overestimate the diffusion of the wave packet.  As an
example, consider a state where the wave packet is 
localized at two vertices separated by a distance $O(N)$. 
Although the state is well localized in the sense that
the particle can be detected only at a few number of sites, 
the variance has a very large value $O(N^2)$. 
This is in contrast with a local network, where the particle can
hop to only locally connected nearby sites and thus $\sigma^2$
may be used as a measure of the diffusion without any confusion. 
To avoid this difficulty with $\sigma^2$, we instead consider 
the participation ratio $P_Q(t)$ as a function of time:
\begin{equation} \label{eq:p}
P_Q(t) \equiv \frac{\sum_n |\Psi_n(t)|^2}{\sum_n |\Psi_n(t)|^4} ,
\end{equation}
which has the value unity for a wave packet localized completely at
one site.  For a completely extended state, on the other hand, 
we have $|\Psi_n|^2 \sim 1/N$ and accordingly, the participation ratio 
$P_Q \sim N$.
The initial localized state with $P_Q(t{=}0) = 1$ thus evolves to
the extended state with $P_Q(t {\rightarrow} \infty) = O(N)$ as the
time-dependent Schr\"odinger equation is integrated in time.

With Eq.~(\ref{eq:Jn}) and the asymptotic form of the Bessel function,
a lengthy but straightforward calculation leads to the asymptotic 
behavior $P_Q (t) \sim t$ on the 1D regular network (without shortcuts)
regardless of the range.
On a $d$-dimensional (regular) network, Eq.~(\ref{eq:H}) still allows the
analytic solution 
$|\Psi_{n_1 n_2 \cdots n_d}(t)| = |J_{n_1}(2t) \cdots J_{n_d}(2t)|$,
producing the behavior $P_Q (t) \sim t^d$.
To characterize diffusion, we define the diffusion time 
$\tau$ associated with the participation ratio by the condition
\begin{equation} \label{eq:tau}
P_Q(t {=} \tau ) =  c N 
\end{equation}
with a constant $c$ between zero and unity.
For a $d$-dimensional network, we thus have the scaling behavior
\begin{equation} \label{ddim}
\tau \sim N^{1/d}.
\end{equation}
For the WS network with shortcuts,
the participation ratio in Eq.~(\ref{eq:p}) is computed numerically 
at given time $t$ and the diffusion time $\tau$ is measured from the condition
in Eq.~(\ref{eq:tau}), where we choose the numerical factor $c=0.25$. 
Other choice for the value of $c$ does not make
any qualitative difference in the scaling behavior of $\tau$,
only if it is not too close to zero or unity.

\begin{figure}
\centering{\resizebox*{!}{6.0cm}{\includegraphics{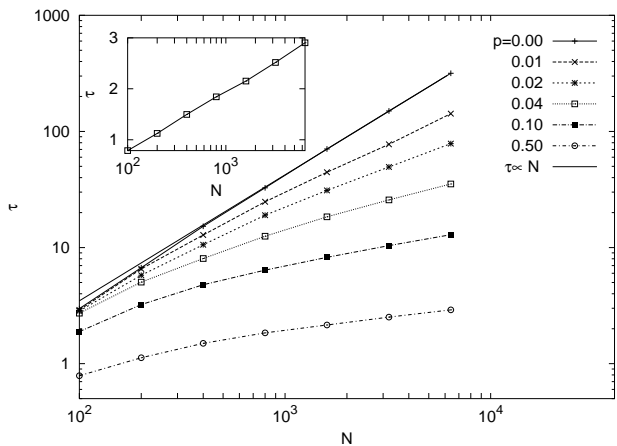}}}
\caption{
Diffusion time $\tau$ 
versus the network size $N$ in the log-log scale.  The case 
$p=0$, corresponding to the local regular network, exhibits 
power-law behavior $\tau \sim N$, 
whereas downward curvatures are present at nonzero values of $p$.
Inset: $\tau$ versus $N$ for $p=0.5$ in the semi-log plot:
The behavior $\tau \sim \log N$ is manifested. Numerical errors
are about the sizes of symbols.
}
\label{fig:QtaupN}
\end{figure}

In Fig.~\ref{fig:QtaupN}, we display the diffusion time $\tau$ 
depending on the size $N$ at various values of the rewiring probability $p$. 
Here the participation ratio $P_Q (t)$ has been computed for 
3000 different network realizations with given parameters $N$, $r$, and $p$,
over which averages have been taken. 
It is observed that the dependence on $N$ changes crucially 
as $p$ is increased: At $p=0$, the linear behavior $\tau \sim N$
is indeed observed in agreement with the analytical result
while the inset manifests the dependence $\tau \sim \log N$ for $p=0.5$. 
We have not systematically investigated the scaling behavior at 
smaller values of $p$; this is difficult from the computational point of view
since $N$ needs to be increased much to avoid finite-size effects. 
Nevertheless it is very plausible to conclude
that the behavior $\tau \sim \log N$ persists as far as $p$ is nonzero. 
The observed change of the scaling behavior 
from $\tau \sim N$ to $\log N$ is very interesting 
in comparison with the change of the characteristic path
length from $l \sim N$ to $\log N$ already observed.~\cite{ref:WS} 
This suggests that the small-world transition at $p=0$, separating the 
large-world behavior ($l \sim N$) from the small-world behavior 
($l \sim \log N$), is also the fast-world transition between 
the slow-world behavior ($\tau \sim N$) and the fast-world behavior 
($\tau \sim \log N$) from the dynamical point of view.
Note also that, in view of Eq. (\ref{ddim}) for
a $d$-dimensional system, such logarithmic behavior apparently indicates that 
the effective dimension of the WS network is infinite, 
which is consistent with the observation of the mean-field 
nature.~\cite{ref:smallIsing,ref:XY}

\begin{figure}
\centering{\resizebox*{!}{6.0cm}{\includegraphics{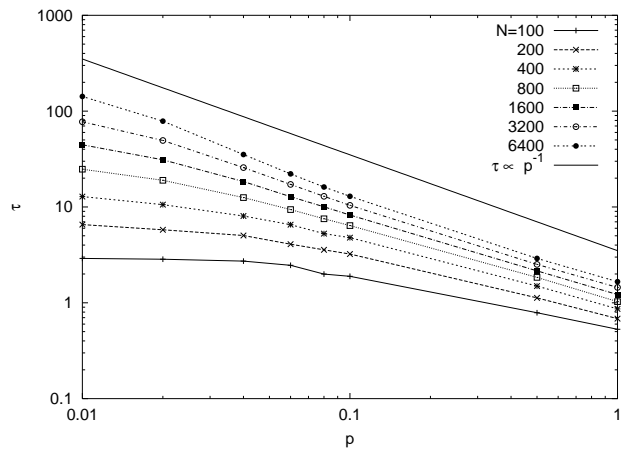}}}
\caption{
Diffusion time $\tau$ versus the rewiring probability $p$
in the WS network of various sizes. 
The behavior $\tau \sim p^{-1}$ is observed in a range of $p$
near $p=1$, which expands as the size $N$ is increased. 
}
\label{fig:QtaupP}
\end{figure}

In Fig.~\ref{fig:QtaupP}, the diffusion time $\tau$ versus the rewiring 
probability $p$ is displayed for various network sizes.  
As the size $N$ becomes larger, the region described well by the relation
$\tau \sim p^{-1}$ covers a broader range of $p$, suggesting that the
power-law behavior $\tau \sim p^{-1}$ is valid at any nonzero value of $p$ in the
thermodynamic limit.  The monotonic decrease of $\tau$ with $p$ 
is easily understood: The more shortcut end points exist, the faster
the diffusion is, as discussed above.
Such anomalous quantum diffusion, observed in this work for complex networks,
has also been investigated in various quasiperiodic quantum 
systems~\cite{ref:GHE,ref:quasiperiodic} as well as in quantum systems
which have classically chaotic counterparts.~\cite{ref:chaos}

%
For comparison, we also consider briefly diffusion of a classical particle 
on the WS network. 
The particle is put on a randomly chosen vertex of the WS network
constructed as before, then allowed to hop to one of its neighboring
vertices, chosen randomly at each time. 
While the return probability in such classical diffusion was examined,~\cite{ref:diff}
it is appropriate here to consider, 
by analogy with the quantum mechanical participation ratio $P_Q (t)$,
the number of vertices visited by the particle during time $t$, denoted by $P_C (t)$. 
Note that this has the two limiting values: $P_C(t{=}0) = 1$ and 
$P_C(t {\rightarrow} \infty) = N$, which are the same as 
those of $P_Q$.  We thus call $P_C$ the classical participation ratio 
for convenience. 
The diffusion time $\tau$ for the classical diffusion is then determined by 
\begin{equation}
P_C(t{=}\tau) = 0.1 N,
\end{equation}
where the use of numerical values other than 0.1 again does not change
the scaling behavior of $\tau$.
In the absence of shortcuts ($p=0$), the WS network reduces to the simple
one-dimensional local regular network, where it is known~\cite{ref:rw} 
$P_C(t) \sim t^{1/2}$, 
and consequently we have the behavior $\tau \sim N^2$. 

\begin{figure}
\centering{\resizebox*{!}{6.0cm}{\includegraphics{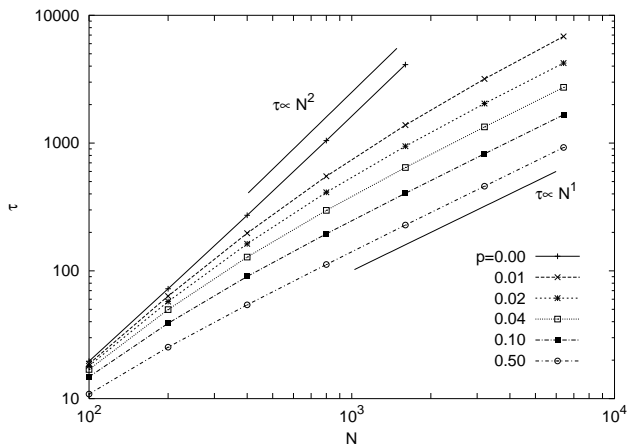}}}
\caption{Diffusion time $\tau$ versus the network size $N$
for classical diffusion. At $p=0$ the behavior $\tau \sim N^2$
is observed, whereas all curves at $p \neq 0$ show the scaling
behavior $\tau \sim N$ for sufficiently large $N$.  
}
\label{fig:CtaupN}
\end{figure}

\begin{figure}
\centering{\resizebox*{!}{6cm}{\includegraphics{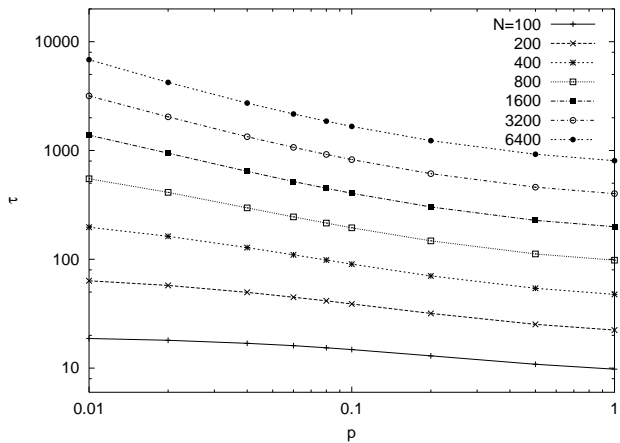}}}
\caption{
Classical diffusion time $\tau$ versus the rewiring probability $p$
for various network sizes.  Unlike quantum diffusion, 
classical diffusion is not well described by the behavior 
of the diffusion time $\tau \sim p^{-1}$. 
}
\label{fig:CtaupP}
\end{figure}

Figure~\ref{fig:CtaupN} shows the diffusion time $\tau$ versus the size $N$ 
of the WS network for various values of the rewiring probability $p$. 
As in the quantum case, $P_C(t)$ has been computed for 3000 different 
network realizations and averages over them have been taken. 
As expected, for $p=0$ corresponding to the local one-dimensional lattice, 
the standard behavior $\tau \sim N^2$ is observed.
In sharp contrast, at any nonzero value of $p$ 
the diffusion time displays the behavior $\tau \sim N$, 
demonstrating that classical diffusion on the WS network changes 
dramatically as soon as $p$ takes a nonzero value. 
This behavior $\tau \sim N$ is also consistent with the 
infinite effective dimension of the WS network,
in view of the known result that $P_C(t) \sim t$ for dimension $d\geq 3$.\cite{ref:rw}
Comparing this observation for classical diffusion with
the previous one for quantum diffusion, 
one can draw in both cases the conclusion that the small-world transition 
at $p=0$, associated with the change of the scaling behavior of the 
characteristic path length $l$, is accompanied by the fast-world transition 
in the dynamic view point, where the exponent $a$ in the relaxation behavior 
$\tau \sim N^a$ changes to $a-1$. 
On the other hand, Fig.~\ref{fig:CtaupP}, where $\tau$ versus $p$ for classical 
diffusion is plotted, shows that the behavior $\tau \sim p^{-1}$ observed for 
quantum diffusion (see Fig.~\ref{fig:QtaupP}) does not appear in classical 
diffusion.

\begin{table}
\label{tab:tau}
\caption{Quantum and classical diffusion: the diffusion time $\tau$
versus the network size $N$. For comparison, the scaling behavior
of the characteristic path length $l$ is also presented.}
\vspace{0.3cm}
\begin{tabular}{ccc}
          &  $p=0$            &  $ p \neq 0$  \\ \hline \hline
quantum   &  $\tau \sim N$    &  $\tau \sim \log N$  \\ \hline
classical &  $\tau \sim N^2$  &  $\tau \sim N$   \\ \hline
characteristic path length & $l \sim N$ & $l \sim \log N$ \\ \hline
\end{tabular}
\label{tab:summary}
\end{table}

%
In summary, we have investigated both quantum diffusion and classical diffusion 
on Watts-Strogatz small-world networks.  To describe these, we have
introduced the participation ratio and determine the diffusion time. 
The obtained scaling behaviors of the diffusion time, 
as summarized in Table~\ref{tab:summary}, show dramatic changes as
one increases the rewiring probability from zero to a nonzero value. 
This is reminiscent of the small-world transition in the scaling behavior
of the characteristic path length $l$ from the small-world regime $l \sim N$
to the large-world regime $l \sim \log N$, and suggests that
the small-world transition can also be termed the fast-world transition.

%
We thank H. Park for useful discussions and acknowledge the partial support
from the Korea Science and Engineering Foundation through Grant 
No.\ R14-2002-062-01000-0 (BJK) and through the SKOREA program 
as well as from the Ministry of Education of Korea through the BK21 project (MYC). 
Numerical simulations have been performed
on the computer cluster Iceberg at Ajou University.

\end{document}